\documentclass[a4paper]{jpconf}

\usepackage{graphicx}

\begin{document}
\title{Irreversibility in a unitary finite-rate protocol: The concept of internal friction}

\author{Sel\c{c}uk \c{C}akmak}
\address{Department of Physics, Ko\c{c} University, \.{I}stanbul, Sar{\i}yer, 34450, Turkey}
\address{Department of Physics, Ondokuz May{\i}s University, Samsun, 55139, Turkey}

\author{Ferdi Altintas}
\address{Department of Physics, Abant \.{I}zzet Baysal University, Bolu, 14280, Turkey}
\address{Department of Physics, Ko\c{c} University, \.{I}stanbul, Sar{\i}yer, 34450, Turkey}

\author{\"{O}zg\"{u}r E. M\"{u}stecapl{\i}o\u{g}lu}
\address{Department of Physics, Ko\c{c} University, \.{I}stanbul, Sar{\i}yer, 34450, Turkey}
\ead{omustecap@ku.edu.tr}

\begin{abstract}
The concept of internal friction, a fully quantum mechanical phenomena, is investigated in a simple, experimentally accessible quantum system in which a spin-1/2 is driven by a transverse magnetic field in a quantum adiabatic process. The irreversible production of the waste energy due to the quantum friction is quantitatively analyzed in a forward-backward unitary transform of the system Hamiltonian by using the quantum relative entropy between the actual density matrix obtained in a parametric transformation and the one in a reversible adiabatic process. Analyzing the role of total transformation time and the different pulse control schemes on the internal friction reveal the non-monotone character of the internal friction as a function of the total protocol time and the possibility for almost frictionless solutions in finite-time transformations.
\end{abstract}

\section{Introduction}
Recent advances in nanotechnology enable successful fabrication and control of systems in length scales where quantum features and fluctuations are dominant. If we are to use these quantum systems for useful purposes, such as heat to work conversion (i.e, as a quantum heat engine), we have to deal with the limitations imposed by quantum mechanics, such as quantum friction, on the thermodynamical transformations and cycles~\cite{plastina14,alecce15,thomas14,feldmann04,feldmann06,kosloff10,rezek06,rezek10,allahver05,allahver08,deffner10,deffner11,deng13,campo14,vaikun09,wei15,ribeiro16,batalhao15}. Quantum friction is the hallmark of finite-time thermodynamical transformations. The infinitely long lasting ones are the reversible processes, which can drive the quantum systems from an equilibrium state to the another one. Reversible processes are, in general, optimal for the work output and the operational efficiency of the quantum heat engines, but they are in the expense of power output. For better powered quantum engines, one typically requires faster thermodynamical transformations, which are irreversible and drive the system outside of equilibrium states, leading to (unwanted) entropy production.

In the present contribution, we investigate the concept of internal friction, which is the irreversibility in closed quantum systems and arises when a quantum system undergoes non-ideal, finite-time parametric adiabatic transformation~\cite{feldmann04,feldmann06,kosloff10,rezek06,rezek10}. Consider a parametrically driven quantum system with a time-dependent Hamiltonian, $H(t)$. Let the system be prepared initially in an equilibrium state with a heat bath. Then, let the system be detached from the heat bath and undergo a parametric unitary change of its Hamiltonian from an initial $H_i$ to a final value $H_f$ in a time interval, $\tau$. If we differ the average work done on the system in a finite-time, $\left\langle w_{\tau}\right\rangle$, and the one performed in an infinite time, $\left\langle w_{\tau\rightarrow\infty}\right\rangle$, the difference is non-negative and introduces the non-adiabatic work performed on the system by the driving agent, i.e., $\left\langle w_{fric}\right\rangle=\left\langle w_{\tau}\right\rangle-\left\langle w_{\tau\rightarrow\infty}\right\rangle \geq 0$~\cite{plastina14}. This indeed defines the internal friction in the system. The system state adiabatically follows the path of equilibrium states in an infinitely long process. Therefore, $\left\langle w_{fric}\right\rangle$ can be considered as a quantitative measure from the deviation of adiabaticity. If the final density matrix for the reversible adiabatic process $(\tau\rightarrow\infty)$ has a well-defined temperature, for example $\beta^{-1}$, then the irreversible work is directly related to the quantum relative entropy between the relevant states through the relation, $<w_{fric}>=\beta^{-1}S(\rho_{\tau}||\rho_{\tau\rightarrow\infty})$~\cite{plastina14}. 

The origin of internal friction is entirely quantum mechanical, as it arises when the system Hamiltonian at different times does not commute, i.e., $[H(t_1), H(t_2)] \neq 0$. It can be associated with the diabatic transitions between energy eigenbasis of the Hamiltonian. When the transformation lasts in a finite time, the system state is unable to follow the temporary changes in the Hamiltonian and therefore develops coherences in the energy frame. This leads to store additional parasitic internal energy in the system, corresponding to the "waste energy", which should be released to a heat bath if we want to thermalize the system at the temperature $\beta^{-1}$ by a subsequent process~\cite{plastina14}. 

The parametric adiabatic transformations are constituents of the quantum Otto and Carnot cycles. Therefore, the role of internal friction in quantum thermal devices has been devoted a considerable attention, recently~\cite{plastina14,alecce15,thomas14,feldmann04,feldmann06,kosloff10,rezek06,rezek10,allahver05,allahver08,deffner10,deffner11,deng13,campo14}. As expected, the quantum friction is found to limit the performance of the quantum heat/refrigerator devices. The strategies to fight against dissipative effects of the internal friction have also been investigated in the context of quantum thermal engines. These strategies can be collect under the common titles as "shortcuts to adiabaticity"~\cite{deng13,campo14} and "quantum lubrication"~\cite{feldmann06}.

In the present study, we will not discuss the minimizing strategies of the internal friction in quantum systems. Our main concern is the concept of the internal friction and to discuss how it arises in a driven quantum system. We follow the ideas and mathematical tools introduced in Refs.~\cite{plastina14,alecce15} by using a simple, experimentally accessible quantum system via NMR setups~\cite{nmr,nmr2} where the quantum system is a two-level system placed in a transverse time-dependent magnetic field. The Hamiltonian of the system is subject to a parametric change from an initial value $H_i$ to a final value $H_f$ in a unitary process (named as forward protocol) followed by a reverse unitary transformation of the Hamiltonian from $H_f$ back to its initial value $H_i$ (named as backward protocol). Due to the misalignment in the magnetic fields of the system, the internal friction  naturally arises. We quantitatively study the deviation from the adiabaticity by using the quantum relative entropy between the initial density matrix of the forward protocol and the final density matrix of the backward protocol. We investigate how the internal friction depends on the transformation time as well as the path that the parametric transformation follows. Our results reveal the non-monotone character of the non-adiabatic work as a function of total protocol time. We also show that considering different possible scheduling pulses that generates the unitary transformation, almost frictionless protocol can be obtained in finite times.

Our results can be relevant to the relation between the macroscopic and microscopic irreversibility (Loschmidt paradox)~\cite{lucia15a} and related time symmetry arguments~\cite{Lucia16b}. Developments in nano-engineering systems makes the question of irreversibility in nano-scale thermodynamics a practical concern~\cite{lucia15b,lucia15c}. Moreover, they can serve for solid examples to clarify and shed some light onto discussions among the novel perspectives of quantum thermodynamics such as unified quantum formulation of mechanics and thermodynamics~\cite{beretta86,maddox85,hatsopoulos76}. In addition, extension of our approach to non-equilibrium thermodynamical systems can be an intriguing application~\cite{lucia16a,nicolis79,beretta06}. We emphasize that dissipative quantum dynamics, entropy production and irreversibility has been widely discussed from different point of views~\cite{korsch87,hensel92,prigogine77,misra79,thoedosopulu78,courbage83}; including nonlinear systems~\cite{svirschevski06}, general description of friction in quantum mechanics with stochastic master equation treatments can be found in the literature, too~\cite{caldirola82}. Our particular case here is limited to irreversibility associated with finite time adiabatic transformations which is the source of a profound quantum internal friction effect when the adiabatic control process is incompatible with the free evolution.

\section{Irreversibility and internal friction}
We consider a driven quantum system which undergoes unitary evolution developed by a parametric time-dependent Hamiltonian of the form,
\begin{equation}\label{eq:ssh}
H(t)=B_0 I_z + B(t) I_x,
\end{equation}
where $B_0$ is a static magnetic field in $z$-direction, while $B(t)$ is a time-dependent magnetic field along the $x$-axes, and $I_{\alpha}$ ($\alpha=x,y,z$) are the components of the spin angular momentum which obey the canonical commutation relation, $[I_{\alpha}, I_{\beta}]=i\epsilon_{\alpha \beta \gamma} I_{\gamma}$. For simplicity, we consider a spin-${1/2}$ particle where $I_{\alpha}=\sigma_{\alpha}/2$ ($\sigma_{\alpha}$ are the Pauli matrices). Throughout the paper, we use a unit system where the Planck constant $\hbar$, the gyromagnetic ratio $\gamma_{n}$, and the Boltzmann constant $k_{B}$ are all set to unity.

The proposed unitary parametric transformation of the Hamiltonian Eq.~(\ref{eq:ssh}) has two independent parts. The first unitary transformation is named as the forward protocol and includes the transformation of the Hamiltonian from $H_i$ to $H_f$ in a finite-time. The latter protocol is the backward transformation of the Hamiltonian from $H_f$ back to $H_i$. The main consequence of such a protocol is to quantitatively measure the internal friction using the initial density matrix of the forward protocol and the final density matrix of the backward protocol, which have the same Hamiltonian $H_i$.

\begin{figure}
\begin{center}
\includegraphics[scale=0.50]{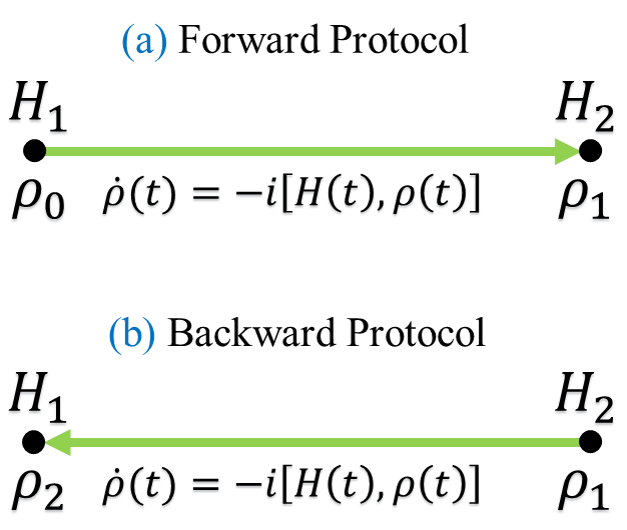}
\caption{\label{fig:0}(Color online.) The schematic diagram of the forward unitary protocol (a) followed by the backward unitary protocol (b). Here $\rho_{\alpha}$ ($\alpha=0,1,2$) and $H_{\alpha}$ ($\alpha=1,2$) are the density matrix and Hamiltonian at the terminal points of the protocols. One should remark that the protocols (a) and (b) are named according to the change of the Hamiltonian. Here, the backward protocol is simply an application a rise-fall protocol for forward time.}
\end{center}
\end{figure}

The schematic representation of the proposed unitary forward-backward protocol is given in Fig.~\ref{fig:0}. The details and the mathematical  formulation are as follow. \textit{(a) Forward Protocol.} The system with initial Hamiltonian $H_1=B_0 I_z + B_1 I_x$ is prepared in an equilibrium with a heat bath at an inverse temperature $\beta=T^{-1}$. The initial density matrix of the system can be given by the Boltzmann-Gibbs distribution $\rho_0=\exp(-\beta H_1)/Z$, where $Z=tr[\exp(-\beta H_1)]$. The system is detached from the heat bath and undergoes an adiabatic parameter change of the magnetic field $B(t)$ from $B_1$ to $B_2$ in a time interval $\tau/2$. The forward protocol can be defined by a unitary evolution of the density matrix given by Liouville-von Neumann equation $\dot{\rho}(t)=-i[H(t),\rho(t)]$, where $H(t)$ is given in Eq.~(\ref{eq:ssh}) and the initial condition is $\rho(t=0)=\rho_0$. The Hamiltonian $H_1$ at $t=0$ is changed to $H_2=B_0 I_z + B_2 I_x$ at the end of the protocol \textit{(a)} with the final density matrix $\rho_1$. There are many possible choices for the scheduling function $B(t)$ to generate this transformation. In the present study, we propose four different driving pulses having explicit forms as $B(t)=B_1 + (B_2-B_1)sin(\pi t/\tau)$ or $B(t)=B_1+(B_2-B_1)(2t/\tau)^n$ (here $n=1/2,1,2$). In all  cases, we have $B(t=0)=B_1$ and $B(t=\tau/2)=B_2$. \textit{(b) Backward Protocol.} This unitary process defines the transformation of the Hamiltonian $H_2$ back to its initial value $H_1$. The time-evolution of the density matrix can be again given by $\dot{\rho}(t)=-i[H(t),\rho(t)]$, where now the initial density matrix is $\rho(t=0)=\rho_1$ and $H(t)$ is given in Eq.~(\ref{eq:ssh}) with the time-dependent magnetic fields having explicit forms as $B(t)=B_2+(B_1-B_2)sin(\pi t/\tau)$ or $B(t)=B_2+(B_1-B_2)(2t/\tau)^n$ ($n=1/2,1,2$). Here $B(t=0)=B_2$ and $B(t=\tau/2)=B_1$. The final density matrix of the backward protocol is denoted by $\rho_2$.

Using Eq.~(\ref{eq:ssh}), one can easily show that the commutator $[H(t_1), H(t_2)]=-iB_0\left(B(t_1)-B(t_2)\right)I_y$, which is always different than zero, since the above proposed magnetic fields are always non-uniform (i.e., $B(t_1) \neq B(t_2)$). Therefore, the system density matrix cannot follow the temporary changes in the Hamiltonian for a fast transformation; the density matrix during the protocol deviates from the equilibrium and thus $\rho_0$ and $\rho_2$ become different. The proper quantification between initial equilibrium state $\rho_0$ and the actual final state of the backward protocol $\rho_2$ can be used as a quantitative measure for the irreversibility in the system~\cite{plastina14,alecce15}. We study the "closeness" of $\rho_2$ to $\rho_0$ through the entropy-like distance measure, namely the quantum relative entropy, defined as~\cite{nielsen10}
\begin{equation}\label{eq:rent0}
S(\rho_2||\rho_0)=tr[\rho_2\ln\rho_2-\rho_2\ln\rho_0]\geq 0.
\end{equation}
According to the Klein's inequality~\cite{nielsen10}, it is non-negative and equals to zero provided that $\rho_2=\rho_0$. Technically, it is not a metric, since it is generally not symmetric, i.e., $S(\rho || \sigma)\neq S(\sigma || \rho)$. The quantum relative entropy has many interpretations in quantum information and computation theory. We refer the readers to recent review~\cite{vedral02} for more details. Remark that if we replace $\rho_2$ by $\rho_0$ in the second term of the right-hand-site of Eq.~(\ref{eq:rent0}), then it will correspond to the difference between von Neumann entropies ($S(\rho)=-tr(\rho\ln\rho)$) of $\rho_2$ and $\rho_0$ which would be zero since von Neumann entropy remains invariant in a unitary path.

We would also like to mention about the calculation procedure of the quantum relative entropy. Eq.~(\ref{eq:rent0}) can be directly evaluated using an appropriate computer software which can handle the logarithm of a matrix. Another calculation for Eq.~(\ref{eq:rent0}) can be done by using the orthonormal decomposition of the density matrices. Denoting $\rho_2=\sum_{i}p_i\left|\psi_i\right\rangle\left\langle \psi_i\right|$ and $\rho_0=\sum_{j}q_j\left|\phi_j\right\rangle\left\langle \phi_j\right|$, the "ln" of the density matrices become $\ln\rho_2=\sum_{i}(\ln p_i)\left|\psi_i\right\rangle\left\langle \psi_i\right|$ and $\ln\rho_0=\sum_{j}(\ln q_j)\left|\phi_j\right\rangle\left\langle \phi_j\right|$. After some calculations, Eq.~(\ref{eq:rent0}) can be rewritten in an equivalent form which is more practical for handmade calculations,
\begin{equation}\label{eq:rent1}
S(\rho_2 || \rho_0)=\sum_{i}p_i\ln p_i-\sum_{i,j}p_i\ln q_j \left|\left\langle \psi_i|\phi_j\right\rangle\right|^2.
\end{equation}

On the other hand, Eq.~(\ref{eq:rent0}) has a deep thermodynamical interpretation; it is intimately related to the internal friction~\cite{plastina14,alecce15}. To make the paper self-consistent, let us prove this assertion here. First, we consider the work performed during the forward-backward protocol in Fig.~\ref{fig:0}. For a reversible adiabatic process ($\tau\rightarrow\infty$), the work performed on the system is zero, since we return to the initial state $\rho_0$ and there would be no change in the internal energy of the system. In this case, the work done on the system by the driving agent in a finite time $\tau$ would be equal to the irreversible work, i.e., $\left\langle w_{fric}\right\rangle=tr[H_1\rho_2]-tr[H_1\rho_0]$. If we consider the thermalization of the state $\rho_2$ to $\rho_0$ during an additional isochoric stage, the average heat supplied to the system by the heat bath at temperature $\beta^{-1}$ would be equal to $\left\langle q_{\rho_2 \rightarrow \rho_0}\right\rangle=-\left\langle w_{fric}\right\rangle$. Now we consider the relation between Eq.~(\ref{eq:rent0}) and the internal friction. Since the von Neumann entropy remains invariant during a unitary evolution, it can be written as $S(\rho_2||\rho_0)=-S(\rho_0)-tr[\rho_2\ln\rho_0]$. Using the above spectral decomposition of $\rho_0$ with $q_j=e^{-\beta E_j}/Z$ where $E_j$ are the eigenvalues of $H_1$, then following some calculation steps, we have $S(\rho_2||\rho_0)=\sum_j \ln q_j \left[ q_j-\left\langle \phi_j\right| \rho_2 \left| \phi_j\right\rangle \right]$. If we distribute $\ln e^{-\beta E_j}/Z=-\beta E_j-\ln Z$ over the sum and use the trace property, we can simplify it as $S(\rho_2||\rho_0)=\beta\left(tr[H_1\rho_2]-tr[H_1\rho_0]\right)$. As a result, the following relation directly holds:
\begin{equation}\label{eq:rfric}
S(\rho_2 || \rho_0)= \beta \left\langle w_{fric}\right\rangle=-\beta\left\langle q_{\rho_2 \rightarrow \rho_0}\right\rangle.
\end{equation}
With respect to Klein's inequality~\cite{nielsen10}, the irreversible work is non-negative; it is stored in the system as a parasitic internal energy (waste energy), as it should have to be released to the heat bath if we want to thermalize the system at the temperature $\beta^{-1}$ with a subsequent thermalization process. We should stress here that the validity of the Eq.~(\ref{eq:rfric}) requires a strong assumption that the developed coherences could be stored in perfect isolation during the entire finite time adiabatic process.
\begin{figure}
\begin{center}
\includegraphics[scale=0.22]{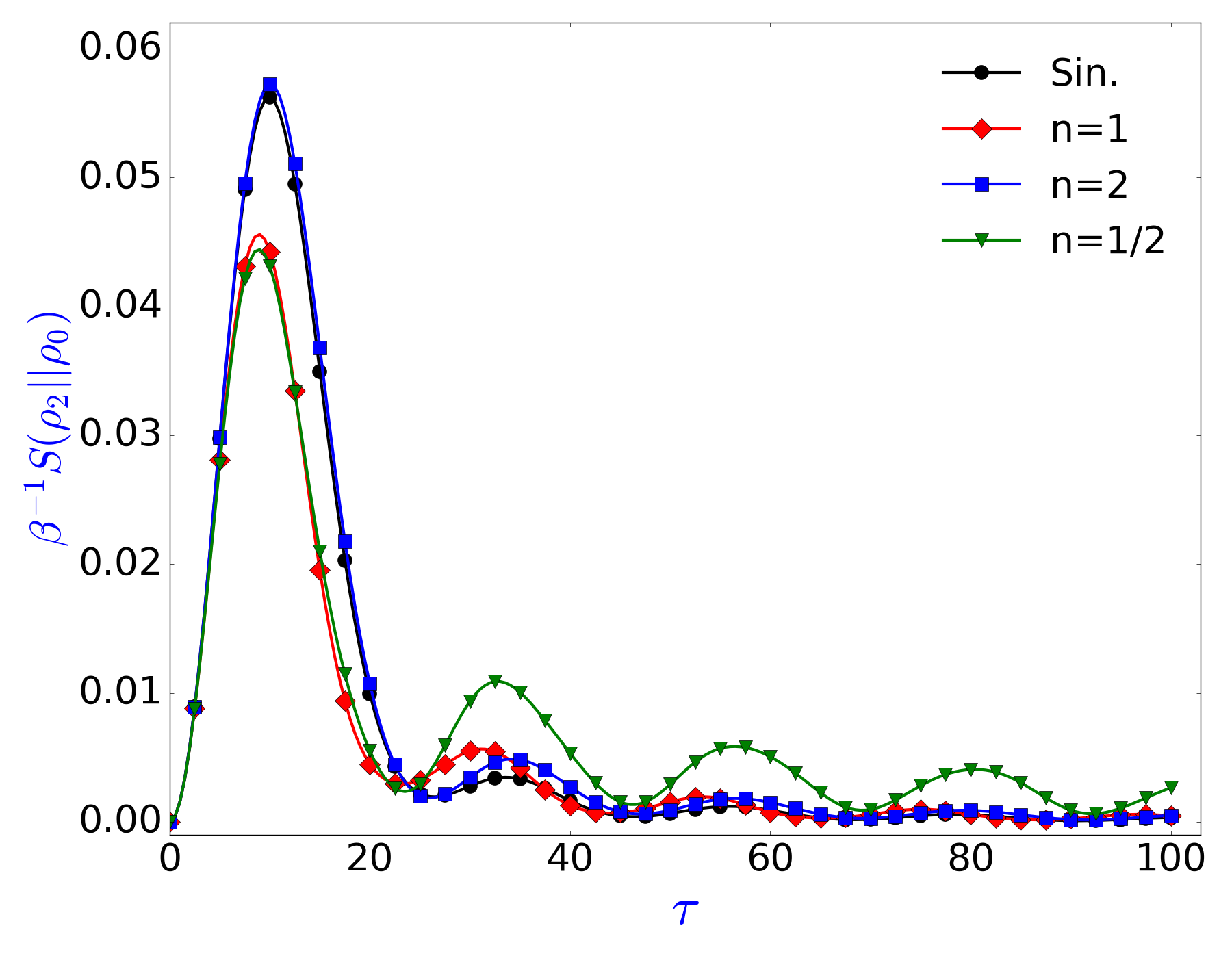}
\caption{\label{fig:1}(Color online.) The quantum relative entropy between the density matrix at the end of the backward protocol $\rho_2$ and the initial thermal state $\rho_0$, $S(\rho_2 || \rho_0)$, as a function of total allocated protocol time $\tau$ for the parameters $B_0=B_1=0.5$, $B_2=0.05$ and $\beta=1$. The unitary transformations are assumed to be generated by four different driving pulses as discussed in the text. Note that the thermodynamical interpretation of the plotted quantity is given in Eq.~(\ref{eq:rfric}).} 
\end{center}
\end{figure}

The parameter $\tau$ defines the total transformation time, as we allot equal time intervals to both protocols. Our main investigation is to analyze how the quantum relative entropy between the state $\rho_2$ and the initial thermal state $\rho_0$ depends on the total protocol time $\tau$. In Fig.~\ref{fig:1}, we plot $S(\rho_2||\rho_0)$ as a function of $\tau$ for the parameters $B_0=B_1=0.5$ and  $B_2=0.05$. Each line in the figure corresponds to the different driving pulses that generates the unitary evolution. Fig.~\ref{fig:1} shows that the quantum relative entropy is non-zero, signifying the irreversible nature of the finite-time transformations; for $S(\rho_2||\rho_0)>0$, the system state cannot return back to its initial state after the backward transformation even when the Hamiltonian can do. Fig.~\ref{fig:1} is a demonstrative example that non-ideal finite-time quantum evolutions are the root of irreversibilities in closed quantum systems.

On the other hand, in the two extreme cases of the total protocol time, we have $\rho_2=\rho_0$. The first limiting case, as shortly discussed above, is the infinitely slow transformation ($\tau \rightarrow \infty$). In this case, the system state follows the temporary changes in the instantaneous Hamiltonian without changing the initial level populations and remains diagonal in terms of the eigenstate representation. Consequently, at the end of the backward protocol, the system state returns to $\rho_0$. This is the condition for the quantum adiabatic theorem to hold in the system and provides frictionless solutions. We should stress here that the transformation timescale should be smaller than the internal timescale of the system, otherwise decoherence effects due to the spontaneous emission of the atoms can take place. The second limiting case is the completely diabatic case ($\tau \rightarrow 0$). In this sudden transformation case, even if the Hamiltonian is transformed, the system density matrix cannot find a chance to be transformed, so it remains unchanged. Therefore, at the end of the backward protocol we will have $\rho_2=\rho_0$ in the limit $\tau \rightarrow 0$. In this scenario, a large variance in energy may take place due to the energy-time uncertainty relation.
\begin{figure}
\begin{center}
\includegraphics[scale=0.15]{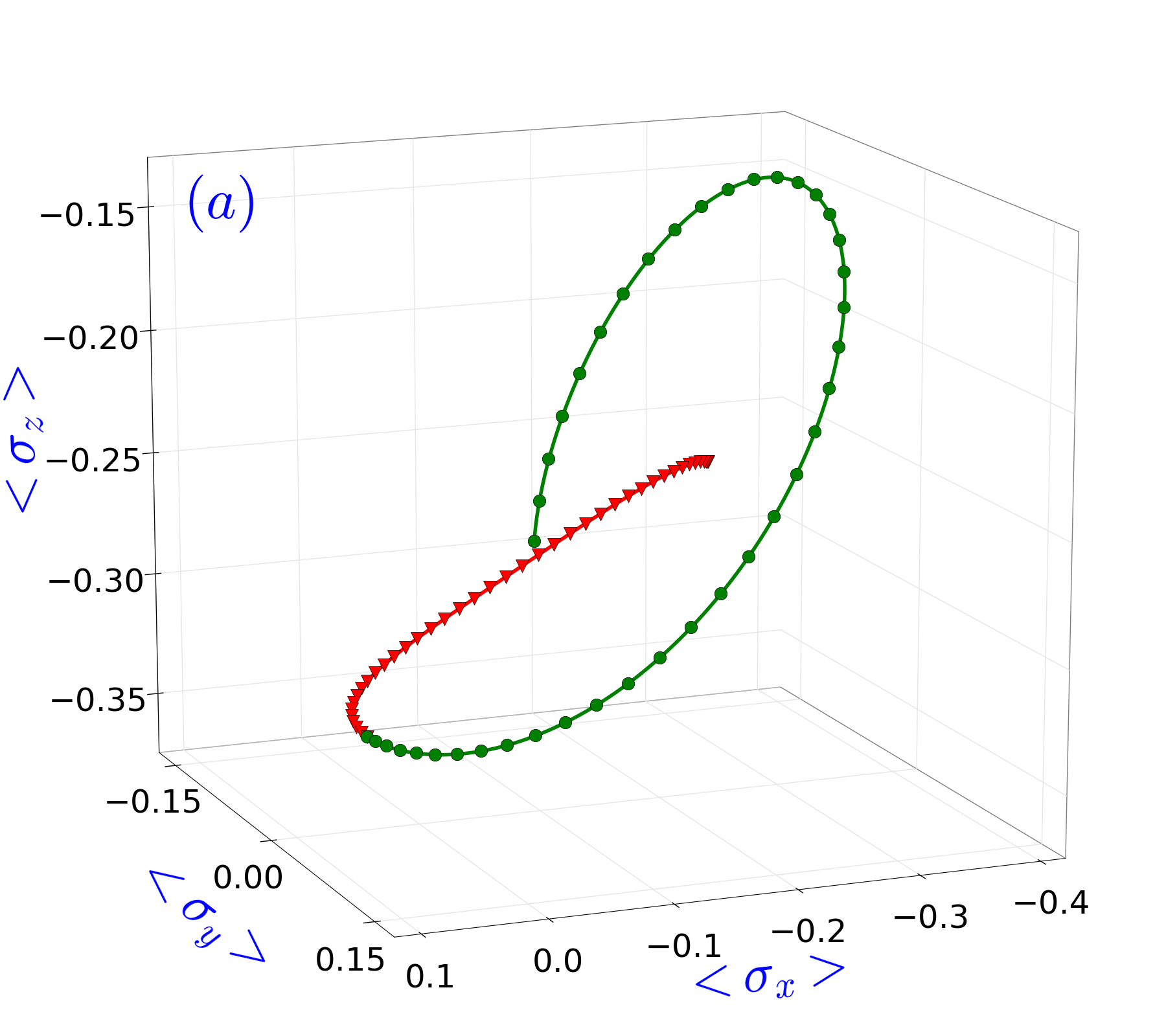}
\includegraphics[scale=0.15]{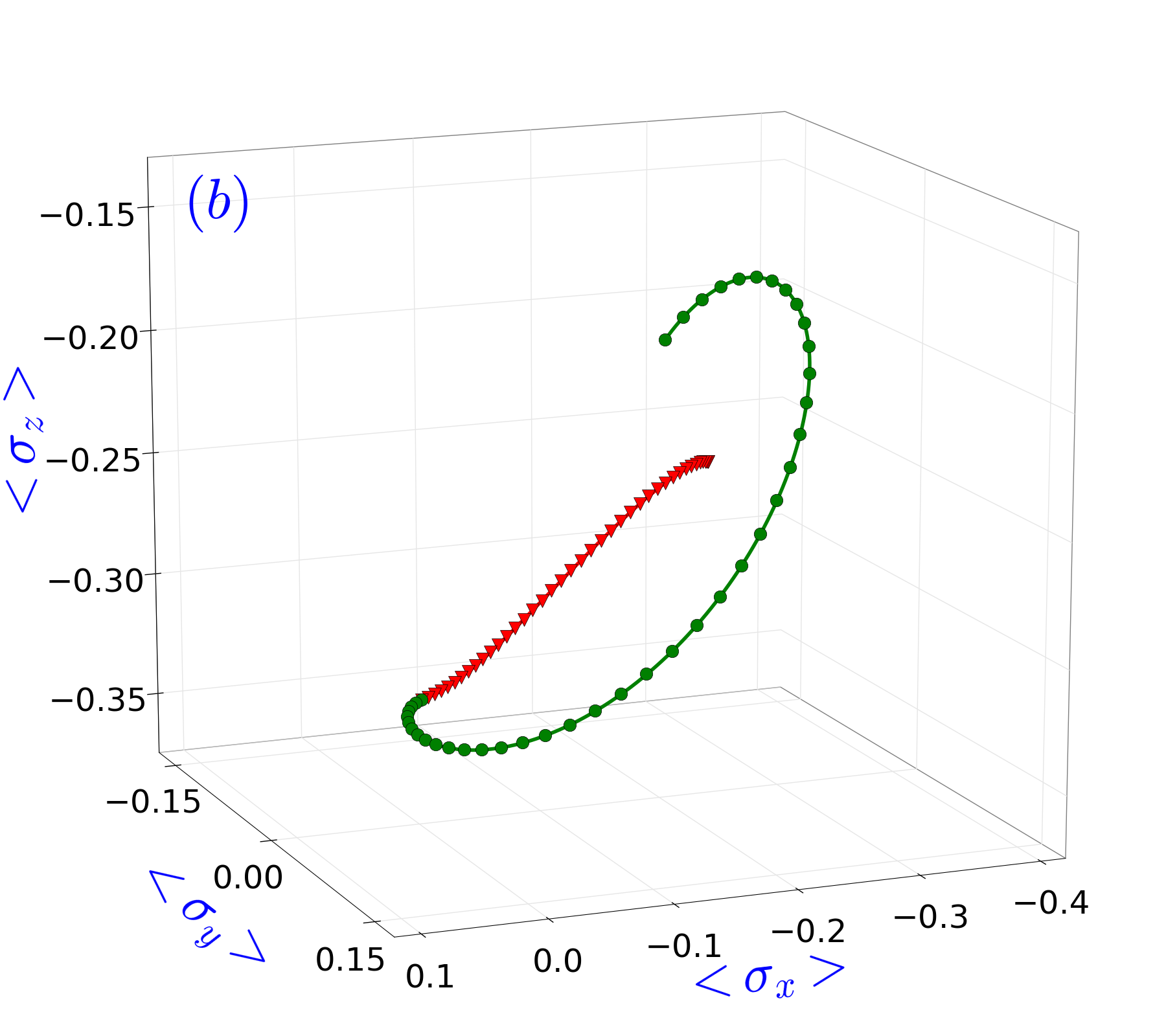}
\includegraphics[scale=0.15]{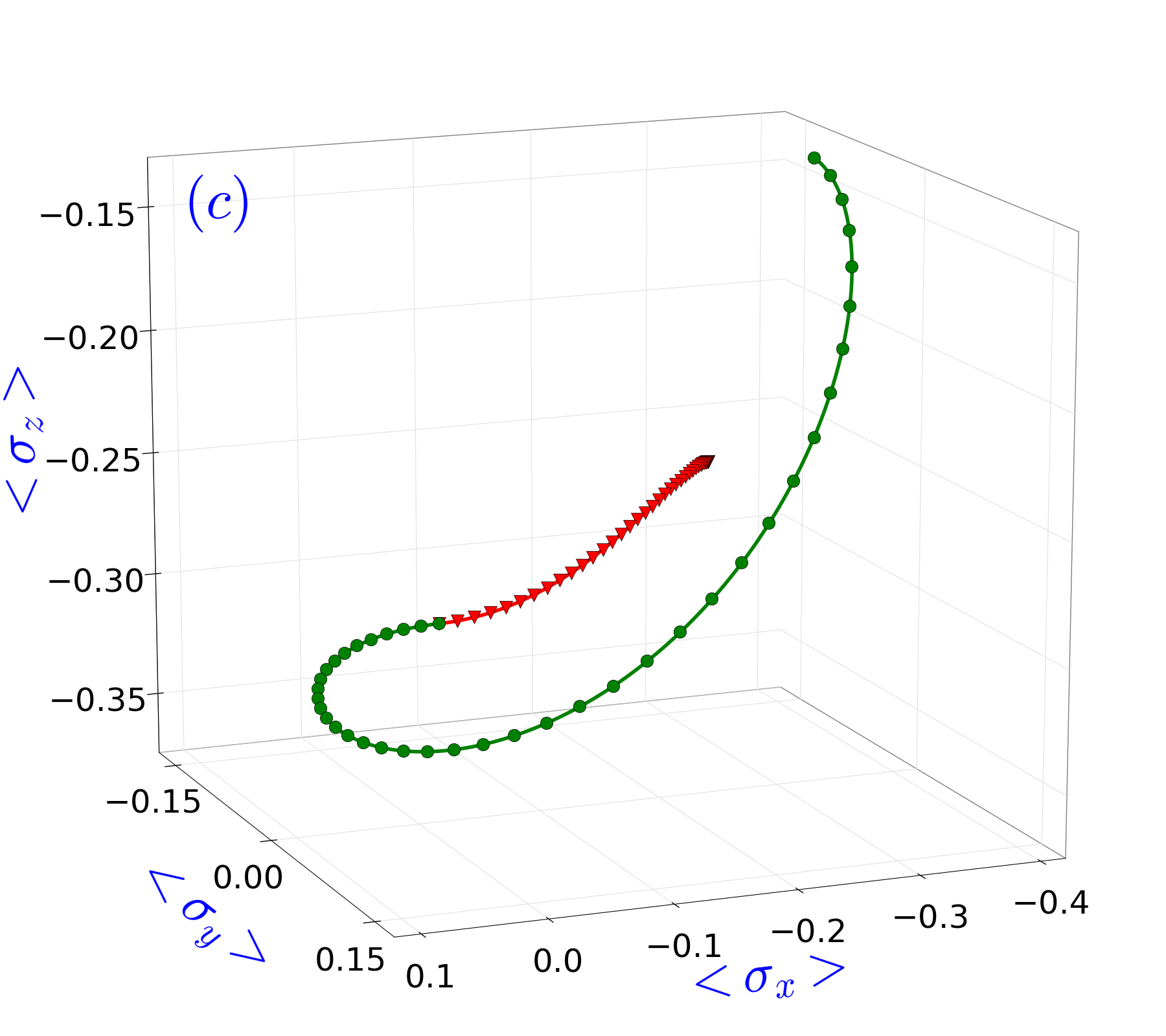}
\includegraphics[scale=0.15]{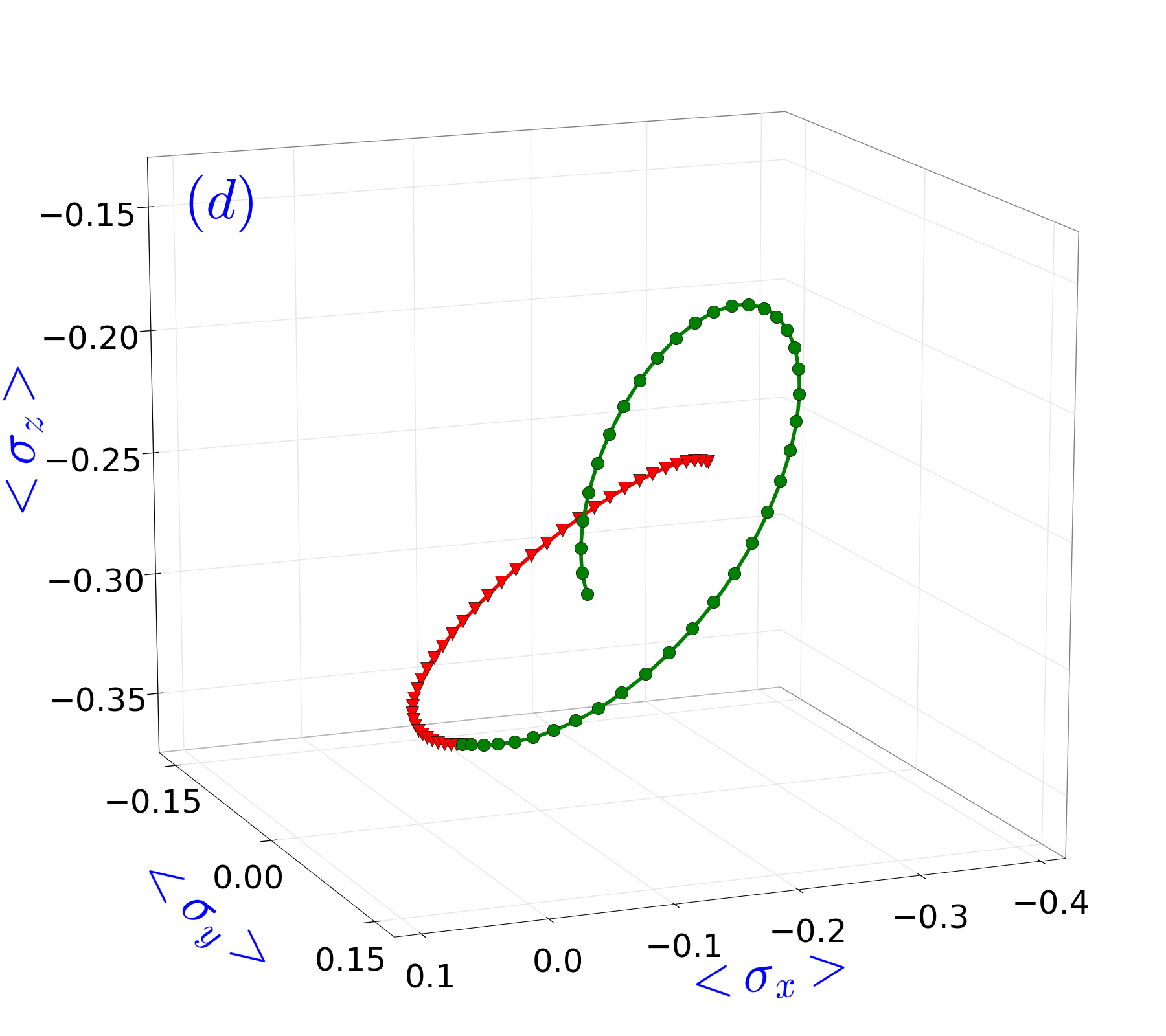}
\caption{\label{fig:2}(Color online.) The trajectory of the Bloch vector in the forward-backward protocol for the parameters given in Fig.~\ref{fig:1} and $\tau=20$. The red, triangular line corresponds to time evolution in the forward protocol, while the green, circle line shows the trajectory in the backward protocol. The sub-figures (a)-(d) are plotted for the driving pulses labeled by sinusoidal, $n=1,2,1/2$, respectively. Remark that the Bloch vector indicates the points in the Bloch sphere corresponding to mixed states. In the forward-backward unitary protocol, the length of the Bloch vector remains intact.} 
\end{center}
\end{figure}

\begin{figure}
\begin{center}
\includegraphics[scale=0.22]{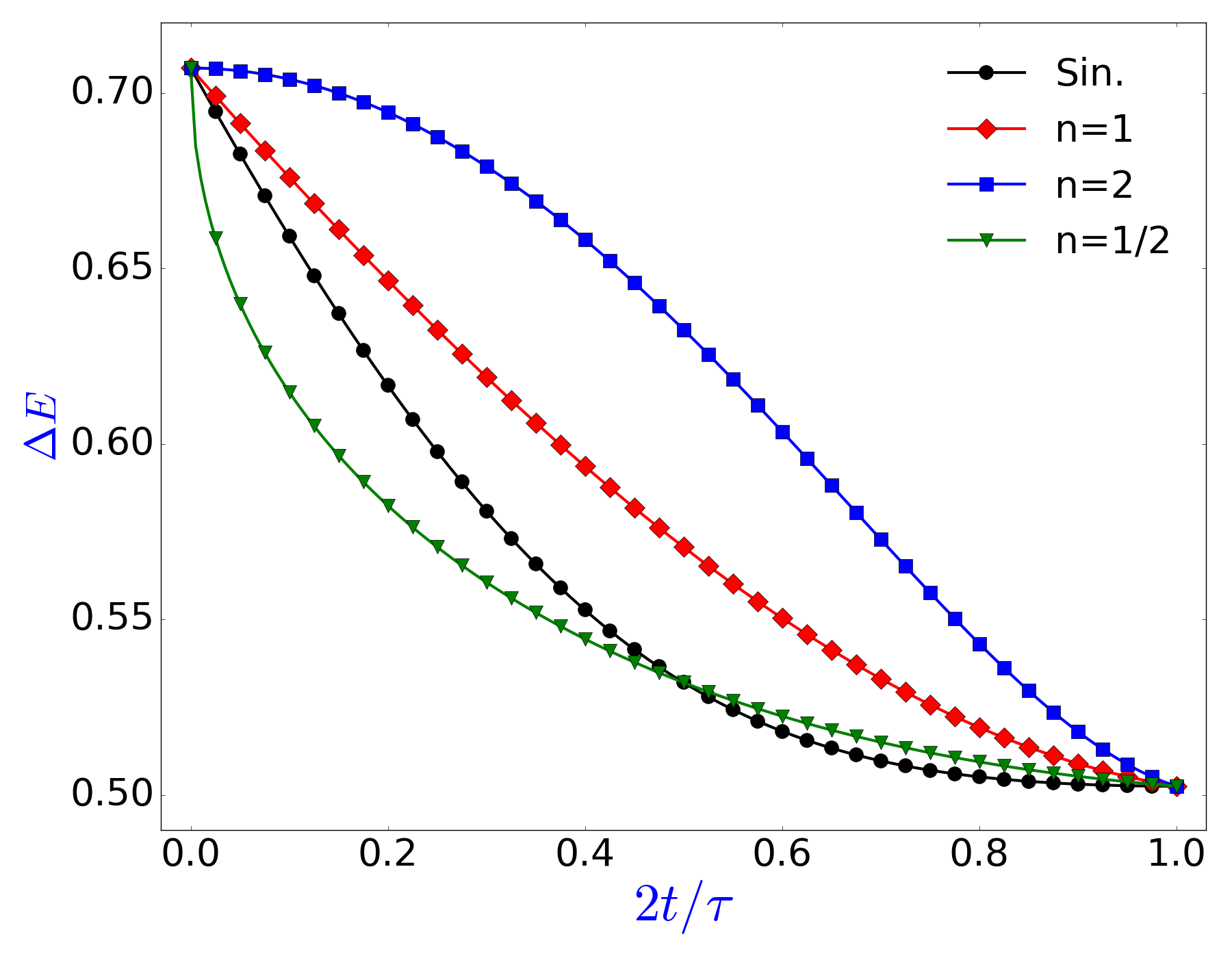}
\caption{\label{fig:3}(Color online.) The time evolution of the energy gap $\Delta E=\sqrt{B_0^2+B(t)^2}$ as a function of $2t/\tau$ in the forward transformation for the parameters given in Fig.~\ref{fig:1} and for the considered driving pulses. The mirror symmetry according to $t=\tau/2$ and the rotational symmetry give the time evolution of the energy gap in the backward protocol.} 
\end{center}
\end{figure}

Fig.~\ref{fig:2} shows the trajectory of the Bloch vector, $\vec{r}=\left\langle \sigma_x \right\rangle \hat{i}+\left\langle \sigma_y \right\rangle \hat{j}+\left\langle \sigma_z \right\rangle \hat{k}$, (i.e., the movement of the density matrix in Bloch
space) in the forward-backward protocol. The Bloch vector cannot return to its initial case due to the quantum friction. Fig.~\ref{fig:3} signifies the time evolution of the energy gap in the protocol. The figure shows that each driving pulse follows different paths for the transformation of the Hamiltonian and defines different protocol rates. This indeed explains the differences in the dynamical behavior of the quantum relative entropy for the different pulses in Fig.~\ref{fig:1} and the trajectory in Fig.~\ref{fig:2}. What is more, the oscillatory behavior of the quantum relative entropy suggests that $S(\rho_2 || \rho_0)$ needs not to depend on the total protocol time monotonically. $S(\rho_2 || \rho_0)$ is maximal around $\tau \approx 10$ for each driving pulse. Fig.~\ref{fig:1} also elucidates that the driving pulses labeled by sinusoidal, $n=1$ and $n=2$ provide "almost" frictionless solutions for shorter protocol time, $\tau>60$, while $n=1/2$ labeled driving pulse requires much longer transformation times. Another interesting observation in Fig.~\ref{fig:1} is the path independent behavior for the almost frictionless solutions around $\tau\approx 20$.

\section{Conclusions}
The concept of internal friction, the irreversibility in a closed quantum system driven out of equilibrium in a unitary transformation, is investigated in a simple, experimentally accessible quantum system. The irreversible production of the excess energy due to the quantum friction is quantitatively analyzed in a forward-backward transform of the Hamiltonian by using the quantum relative entropy between the actual density matrix obtained during the parametric transformation and the one taken in the infinitely slow limit. The role of total transformation time and the different pulse control schemes on the internal friction are investigated in detail. The transformations taken in finite-times are irreversible processes, where the system density matrix cannot return back to its initial state. This is also verified from the trajectory plots of the density matrix in Bloch space. Our proposed pulse control scheme reveals the non-monotone dependence of the internal friction on the total protocol time and the possibility for almost frictionless solutions in finite-time transformations.

\section*{Acknowledgments}
The authors thank Dieter Suter for fruitful discussions. S.\c{C}. warmly thanks Azmi Gen\c{c}ten for useful discussions. F.A. thanks Department of Physics of the Ko\c{c} University for hospitality. The authors acknowledge support from Ko\c{c} University and Lockheed Martin Corporation Research Agreement.


\section*{References}

\begin{thebibliography}{}

\bibitem{plastina14} Plastina F, Alecce A, Apollaro T J G, Falcone G, Francica G, Galve F, Gullo N Lo and Zambrini R 2014 {\it Phys. Rev. Lett.} {\bf 113} 260601

\bibitem{alecce15} Alecce A, Galve F, Gullo N L, Dell'Anna L, Plastina F and Zambrini R 2015 {\it New J. Phys.} {\bf 17} 075007

\bibitem{thomas14} Thomas G and Johal R S 2014 {\it Eur. Phys. J.} B {\bf 87} 166

\bibitem{feldmann04} Feldmann T and Kosloff R 2004 {\it Phys. Rev.} E {\bf 70} 046110

\bibitem{feldmann06} Feldmann T and Kosloff R 2006 {\it Phys. Rev.} E {\bf 73} 025107

\bibitem{kosloff10} Kosloff R and Feldmann T 2010 {\it Phys. Rev.} E {\bf 82} 011134

\bibitem{rezek06} Rezek Y and Kosloff R 2006 {\it New J. Phys.} {\bf 8} 83

\bibitem{rezek10} Rezek Y 2010 {\it Entropy} {\bf 12} 1885

\bibitem{allahver05} Allahverdyan A E and Nieuwenhuizen Th M 2005 {\it Phys. Rev.} E {\bf 71} 046107

\bibitem{allahver08} Allahverdyan A E, Johal R S and Mahler G 2008 {\it Phys. Rev.} E {\bf 77} 041118

\bibitem{deffner10} Deffner S and Lutz E 2010 {\it Phys. Rev. Lett.} {\bf 105} 170402

\bibitem{deffner11} Deffner S and Lutz E 2011 {\it Phys. Rev. Lett.} {\bf 107} 140404

\bibitem{deng13} Deng J, Wang Q-h, Liu Z, Hanggi P and Gong J 2013 {\it Phys. Rev.} E {\bf 88} 062122

\bibitem{campo14} del Campo A, Goold J and Paternostro M 2014 {\it Scientific Reports} {\bf 4} 6208 

\bibitem{vaikun09} Vaikuntanathan S and Jarzynski C 2009 {\it Europhysics Lett.} {\bf 87} 60005

\bibitem{wei15} Wei Bo-Bo and Plenio M B 2015 {\it arXiv:1509.07043}

\bibitem{ribeiro16} Ribeiro W L, Landi G T and Semiao F L 2016 {\it arXiv:1601.01833} 

\bibitem{batalhao15} Batalhao T B, Souza A M, Sarthour R S, Oliveira I S, Paternostro M, Lutz E and Serra R M 2015 {\it Phys. Rev. Lett.} {\bf 115} 190601

\bibitem{nmr} Stolze J and Suter D 2008 {\it Quantum Computing: A Short Course from Theory to Experiment} (Wiley)

\bibitem{nmr2} Batalhao T B, Souza A M, Mazzola L, Auccaise R, Sarthour R S, Oliveira I S, Goold J, Chiara G D, Paternostro M and Serra R M 2014 {\it Phys. Rev. Lett.} {\bf 113} 140601

\bibitem{lucia15a} Lucia U 2015 {\it Chem. Phys. Lett.} {\bf 623} 98-100

\bibitem{Lucia16b} Lucia U 2016 {\it Physica A} {\bf 444} 121-128

\bibitem{lucia15b} Lucia U 2015 {\it Entropy} {\bf 17}(2) 1309-1328

\bibitem{lucia15c} Lucia U 2015 {\it Chem. Phys. Lett.} {\bf 629} 87-90

\bibitem{beretta86} Beretta G P 1986 {\it Frontiers of Nonequilibrium Statistical Physics: Proc of the NATO Advanced 
Study Institute, (Santa Fe, 1984, Series B: Physics, Vol. 135)} ed G.T. Moore and M.O. Scully 
(New York: Plenum) p 205

\bibitem{maddox85} Maddox J 1985 {\it Nature}. {\bf 316} 11

\bibitem{hatsopoulos76} Hatsopoulos G N and Gyftopoulos E P 1976 {\it Found. Phys.} {\bf 6} 15; 1976 {\it Found. Phys.} {\bf 6} 127; 1976 {\it Found. Phys.} {\bf 6} 439; 1976 {\it Found. Phys.} {\bf 6} 561

\bibitem{lucia16a} Lucia U 2016 {\it Physica A} {\bf 447} 314-319

\bibitem{nicolis79} Nicolis G and Prigogine I 1979 {\it Proc. Natl. Acad. Sci. U.S.} {\bf 76} 6060

\bibitem{beretta06} Beretta G P 2006 {\it Phys. Rev. E} {\bf 73} 026113

\bibitem{korsch87} Korsch H J and Steffen H 1987 {\it J. Phys. A} {\bf 20} 3787

\bibitem{hensel92} Hensel M and Korsch J 1992 {\it J. Phys. A} {\bf 25} 2043

\bibitem{prigogine77} Prigogine I, Mayne F, George C and Haan M De 1977 {\it Proc. Natl. Acad.
Sci. U.S.} {\bf 74} 4152

\bibitem{misra79} Misra B, Prigogine I and Courbage M 1979 {\it Proc. Natl. Acad.
Sci. U.S.} {\bf 76} 4768

\bibitem{thoedosopulu78} Thoedosopulu M, Grecos A and Prigogine I 1978 {\it Proc. Natl. Acad.
Sci. U.S.} {\bf 75} 1632

\bibitem{courbage83} Courbage M and Prigogine I 1983 {\it Proc. Natl. Acad. Sci. U.S.} {\bf 80} 2412

\bibitem{svirschevski06} Svirschevski G P 2001 {\it Phys. Rev. A} {\bf 63} 022105; 2001 {\it Phys. Rev. A} {\bf 63} 054102

\bibitem{caldirola82} Caldirola P and Lugiato L A 1982 {\it Physica A} {\bf 116} 248

\bibitem{nielsen10} Nielsen M A and Chuang I L 2010 {\it Quantum Computation and Quantum Information} (Cambridge: Cambridge University Press)

\bibitem{vedral02} Vedral V 2002 {\it Rev. Mod. Phys.} {\bf 74} 197

\end{thebibliography}
\end{document}